\documentclass[journal,12pt,onecolumn,draftclsnofoot,]{IEEEtran}
\IEEEoverridecommandlockouts
\usepackage{cite}
\usepackage{amsmath,amssymb,amsfonts}
\usepackage{algorithmic}
\usepackage{graphicx}
\usepackage{textcomp}
\usepackage{xcolor}
\usepackage{dsfont}
\usepackage{subcaption}
\usepackage{float}
\usepackage{amsmath}
\usepackage[]{graphicx}
\usepackage{lipsum}

\def\BibTeX{{\rm B\kern-.05em{\sc i\kern-.025em b}\kern-.08em
    T\kern-.1667em\lower.7ex\hbox{E}\kern-.125emX}}

\begin{document}

\title{Coexistence of URLLC and eMBB services in the C-RAN Uplink: An Information-Theoretic Study\\
\thanks{The work of Rahif Kassab, Osvaldo Simeone and Petar Popovski has received funding from the European  Research  Council (ERC) under the European Union Horizon 2020 research and innovation program (grant agreements 725731 and 648382).}
}



\author{\IEEEauthorblockN{Rahif Kassab\IEEEauthorrefmark{1},
Osvaldo Simeone\IEEEauthorrefmark{1} and Petar Popovski\IEEEauthorrefmark{2} }\\
\IEEEauthorblockA{\small\IEEEauthorrefmark{1}Centre for Telecommunications Research, King's College London, London, United Kingdom\\
\IEEEauthorrefmark{2}Department of Electronic Systems, Aalborg University, Aalborg, Denmark\\
Emails: \IEEEauthorrefmark{1}\{rahif.kassab,osvaldo.simeone\}@kcl.ac.uk,
\IEEEauthorrefmark{2}petarp@es.aau.dk}}

\vspace{-10 mm}
\maketitle

\begin{abstract}
The performance of orthogonal and non-orthogonal multiple access is studied for the multiplexing of enhanced Mobile BroadBand (eMBB) and Ultra-Reliable Low-Latency Communications (URLLC) users in the uplink of a multi-cell Cloud Radio Access Network (C-RAN) architecture. While eMBB users can operate over long codewords spread in time and frequency, URLLC users' transmissions are random and localized in time due to their low-latency requirements. These requirements also call for decoding of their packets to be carried out at the edge nodes (ENs), whereas eMBB traffic can leverage the interference management capabilities of centralized decoding at the cloud. Using information-theoretic arguments, the performance trade-offs between eMBB and URLLC traffic types are investigated in terms of rate for the former, and rate, access latency, and reliability for the latter. The analysis includes non-orthogonal multiple access (NOMA) with different decoding architectures, such as puncturing and successive interference cancellation (SIC). The study sheds light into effective design choices as a function of inter-cell interference, signal-to-noise ratio levels, and fronthaul capacity constraints.    
\end{abstract}


\section{Introduction}
The fifth generation (5G) of wireless cellular systems is expected to cater to three generic services, namely Enhanced Mobile BroadBand (eMBB), Ultra-Reliable Low-Latency Communications, and massive Machine-Type Communications (mMTC) \cite{itu}\cite{3gpp}. eMBB traffic allows for higher transmission rates as compared to current (4G) systems, and it can leverage coding over large transmission blocks due to its non-critical latency requirements. In contrast, URLLC imposes strict latency constraints, typically of 0.25-0.3 ms/packet, hence requiring transmissions localized in time, while still ensuring high reliability levels. mMTC traffic consists of a large number of uncoordinated devices transmitting small payloads to a common receiver. \par
The standard approach to guarantee the mentioned heterogeneous quality-of-service requirements for all services is to implement orthogonal multiple access (OMA), or orthogonal slicing, whereby distinct radio resources are reserved for use of eMBB, URLLC and mMTC users (see, e.g., Fig. 1(a)) \cite{5gtutorial}, \cite{urllcpopovski}. However, OMA can be strictly suboptimal, since URLLC and mMTC users’ activities are bursty and generally unpredictable, and hence resources allocated exclusively to such users may be wasted.\par

\begin{figure}[t]
	\centering
	\includegraphics[height= 10 cm, width= 10 cm]{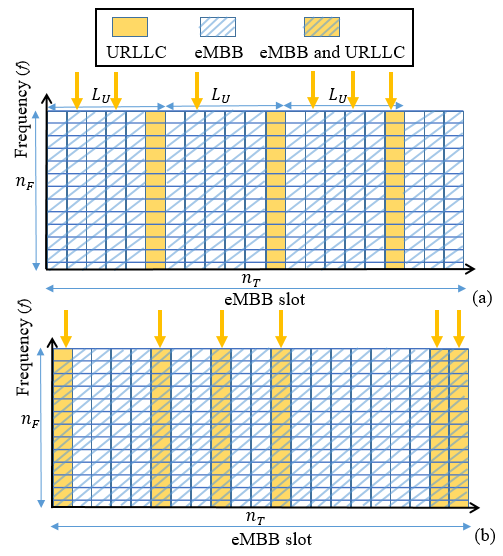}
	\caption{Time-frequency resource allocation for: (a) Orthogonal Multiple Access Scheme (OMA); and (b) Non Orthogonal Access Scheme (NOMA). Hatched areas correspond to eMBB transmissions. Downward arrows denote arrivals of URLLC packets. }
	\label{fig:time_frequency}
    \vspace{-5 mm}
\end{figure}
 
In order to improve the efficiency of OMA, non-orthogonal multiple access (NOMA) techniques enable the simultaneous transmissions of users from different services (see Fig. 1(b)). The coexistence on the same radio resources of eMBB and URLLC users is studied in \cite{Anand2017JointSO} \cite{networkslicingpopvskiandsimeone} for a \textit{single cell}. Specifically, reference \cite{Anand2017JointSO} focuses on the problem of scheduling URLLC users by abstracting the performance at the physical layer, while \cite{networkslicingpopvskiandsimeone} presents a communication-theoretic model. The latter reference also considers the performance of NOMA for eMBB and mMTC devices.
\par
In this paper, we study for the first time the performance of OMA and NOMA for the multiplexing of eMBB and URLLC users in the uplink of a \textit{multi-cell Cloud Radio Access Network} (C-RAN) architecture, as illustrated in Fig. 2. In a C-RAN, Edge Nodes  (ENs) in different cells are connected to a Baseband Unit (BBU) in the cloud by means of fronthaul links \cite{bookcransimeone}. As summarized in Fig. 1, eMBB users can operate over long codewords spanning the time-frequency plane, while URLLC users' transmissions are random and localized in time due to their low-latency requirements. These also call for the decoding of URLLC packets at the ENs (see also \cite{wiggerdelay}), while decoding of eMBB users can leverage the interference management capabilities of the BBU as in the standard C-RAN architecture (see e.g., \cite{bookcransimeone}).\par
We aim at characterizing the performance trade-off between eMBB and URLLC traffic types in terms of rate for the former, and rate, access latency, and reliability for the latter. The analysis includes OMA, as well as NOMA with different decoding architectures, such as puncturing and Successive Interference Cancellation (SIC). This study leverages information-theoretic arguments, and it accounts for inter-cell interference, URLLC access errors due to an insufficient number of transmission opportunities, and fronthaul capacity constraints.
\par 
The rest of the paper is organized as follows. Sec. II describes the system and signal models for both URLLC and eMBB users, as well as the relevant performance metrics. In Sec. III and Sec. IV, we analyze the performance of OMA and NOMA by considering different decoding schemes for the latter, namely puncturing, treating URLLC as noise, and SIC. In Sec. V, numerical results and related discussion are provided, and conclusions are drawn in Sec. VI.

\begin{figure}[t]
	\centering
	\includegraphics[height= 8 cm, width= 10 cm]{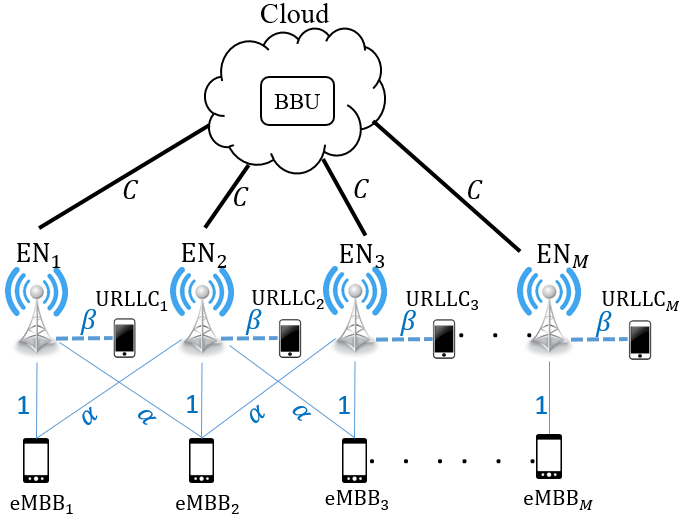}
	\caption{A C-RAN system with coexisting eMBB and URLLC users. }
	\label{fig:system_model}
\end{figure}

\section{System and Signal Model}
As illustrated in Fig.~\ref{fig:system_model}, we consider a generalization of the Wyner model \cite{simeone2012cooperative} that encompasses both eMBB and URLLC users. 
\subsection{System Model}
Cells are arranged in a line, with one EN per cell. Each cell contains two users active in the given radio resources, namely an eMBB and an URLLC user. All ENs are connected to a BBU in the cloud by mean of orthogonal fronthaul links. Focusing on the uplink, we assume that the URLLC users are located close to the ENs, and are hence received with non-negligible power only by the EN in the same cell. This condition may be ensured by scheduling only users close to the EN that can satisfy the high URLLC reliability requirements. Alternatively, in mission-critical or Industry 4.0 scenarios, ENs can be deployed where URLLC devices are expected to be present. The eMBB users, instead, need not to satisfy this condition, and are assumed to be close to the cell boundary in order to focus on worst-case performance guarantees. As a result, the eMBB users are received with non-negligible power by the EN in the same cell and by the ENs in the left and right neighboring cells \cite{simeone2012cooperative}. \par
All channel gains are normalized with respect to the direct channel from an eMBB user to the in-cell EN, which is set to one. Accordingly, the inter-cell eMBB channel gain is equal to $\alpha \in [0,1]$, and the URLLC users have a channel gain $\beta \geq 0$. All channel gains are assumed to be constant in the spectral resources under study shown in Fig.~\ref{fig:time_frequency}. The fronthaul link between each $\mathrm{EN}$ and the cloud has a finite capacity of $C$ bit/s/Hz.\par 
As illustrated in Fig.~\ref{fig:time_frequency}, we assume that the time and frequency plane is divided into radio resources, each occupying one minislot and one frequency channel. We focus on frames of $n_T$ minislots, with each minislot composed of $n_F$ frequency channels. For simplicity, we assume that each radio resource accommodates the transmission of a single symbol, but the analysis can be readily generalized. We will refer to the minislot index as $t \in [1,n_T]$ and to the frequency channel index as $f \in [1,n_F]$.\par 
The eMBB users transmit over the entire time-frequency frame, and hence the maximum transmission blocklength for the eMBB users is equal to $n_{T}n_{F}$. Due to the latency constraints of the URLLC traffic, each URLLC transmission can instead span only a single minislot, and hence the blocklength of URLLC traffic is equal to $n_F$ symbols. URLLC packers are generally small and we have the condition $n_F \ll n_T$. \par 
As illustrated in Fig.~\ref{fig:time_frequency}, each URLLC user generates an independent packet in each minislot with probability $q$. As detailed below, the URLLC packet is transmitted at the next available transmission opportunity if possible.\par
\subsection{Signal Model}
As seen in Fig. 1(a), with OMA, one minislot is exclusively allocated to URLLC users every $L_U$ minislots. If the $k$-th URLLC user generates more than one packet in the $L_U$ minislots between two transmission opportunities, one such packet is selected randomly for transmission and the others are discarded. This guarantees a \textit{worst-case access latency} of $L_U \geq 1$ minislots, with $L_U=1$ indicating the minimal latency. The signal $Y_k^{f}(t)$ received at each $\mathrm{EN}_k$ with $k=1, \ldots , M$ and frequency $f=1,\ldots,n_F$ can hence be written as 
\begin{equation}
  Y_{k}^{f}(t)=\begin{cases}
   \beta A_k(t) U_k^{f}(t) + Z_{k}^{f}(t), & \text{if\ $t=L_U, 2L_U,\ldots$}\\
    X_k^{f}(t) + \alpha X_{[k-1]}^{f}(t) + \alpha X_{[k+1]}^{f}(t) + Z_k^{f}(t), & \text{otherwise},
  \end{cases} \label{eq:1}
\end{equation} where $X_{k}^{f}(t)$ denotes the signal transmitted by the $k$-th eMBB user over subcarrier $f$ at time $t$; $Z_{k}^{f}(t) \sim \mathcal{CN}(0,1)$ is Complex Gaussian noise with zero and unit variance, which is i.i.d. across the indices $k,f$ and $t$; $U_k^{f}(t)$ denotes the URLLC signal sent by the $k$-th URLLC user; and $A_k(t)$ is an indicator variable that equals one if a URLLC user transmits at time $t$ and zero otherwise. In order to ensure symmetry, in \eqref{eq:1} and \eqref{eq:Y_noma}, we consider a circulant Wyner model \cite{simeone2012cooperative}, in which $[k-1]=M$ for $k=1$ and $[k+1]=1$ for $k=M$. \par 
As seen in Fig. 1(b), for NOMA, URLLC users transmit immediately in the minislot in which a packet is generated, so that the access latency is minimal, i.e., $L_U=1$. Accordingly, the signal $Y_{k}^{f}(t)$ received at each $\mathrm{EN}_k$ with $k=1, \ldots , M$ at time $t=1,\ldots,n_T$ and frequency $f=1,\ldots,n_F$ can be written as
\begin{equation}
Y_{k}^{f}(t)=  X_k^{f}(t) + \alpha X_{[k-1]}^{f}(t) + \alpha X_{[k+1]}^{f}(t) +\beta A_k(t) U_k^{f}(t) + Z_k^{f}(t), \label{eq:Y_noma}
\end{equation}
with definitions as in \eqref{eq:1}. \par
The power constraint for eMBB and URLLC users are defined respectively as
\begin{equation}
\frac{1}{n_{T}n_{F}} \sum_{t=1}^{n_T} \sum_{f=1}^{n_F} \mathbb{E}[|X_{k}^{f}(t)|^{2} ]\leq P_B \label{eq:constraint_B}
\end{equation}
\begin{equation}
\mathrm{and}\ \frac{1}{n_{F}} \sum_{f=1}^{n_F} \mathbb{E}[|U_{k}^{f}(t)|^{2}] \leq P_U \label{eq:constraint_C},
\end{equation}
where the temporal average in \eqref{eq:constraint_B}-\eqref{eq:constraint_C} is taken over all symbols within a codeword. We also assume that the channel coefficients $\alpha$ and $\beta$ are known to all users and ENs.\par
Model \eqref{eq:1} and \eqref{eq:Y_noma} can be written in matrix form by introducing the $M \times n_F$ matrix $\mathbf{X}(t)$, whose $(k,f)$ entry is given by $X_{k}^{f}(t)$, and, in a similar manner, the $M \times n_F$ matrices $\mathbf{U}(t)$, $\mathbf{Y}(t)$ and $\mathbf{Z}(t)$.
Defining also, the channel matrix \textbf{H} as an $M\times M$ circulant matrix with first column given by the vector $[1 \ \alpha \ 0 \ldots \  0\ \alpha]^{\mathsf{T}}$, we can write the received signal \eqref{eq:Y_noma} across all ENs as
\begin{equation}
\mathbf{Y}(t) = \mathbf{HX}(t)+\beta \mathbf{A}(t)\mathbf{U}(t) + \mathbf{Z}(t),
\end{equation}
where $\mathbf{A}$ is a diagonal $M\times M$ matrix with $(i,i)$ entry equal to $A(i,i)\sim \mathcal{B}(q)$ for each $i \in [1,M]$. Model \eqref{eq:1} can be written in an analogous way.
We will also find it useful to write the signals received at frequency $f$ across all ENs as
\begin{equation}
\mathbf{y}^{f}(t) = \mathbf{Hx}^{f}(t)+\beta \mathbf{Az}^{f}(t) + \mathbf{z}^{f}(t),
\end{equation}
where $\mathbf{y}^{f}(t), \mathbf{x}^{f}(t),\mathbf{z}^{f}(t)$ and $\mathbf{z}^{f}(t)$ are the $f$-th columns of matrices $\mathbf{Y}(t), \mathbf{X}(t), \mathbf{U}(t)$ and $\mathbf{Z}(t)$, respectively. In the following, we will drop the dependence on $t$ when no confusion may arise.
\subsection{Performance Metrics}
We are interested in the following performance metrics.
For eMBB users, we study the per-cell sum rate
\begin{equation}
R_B=\frac{\log_2 (M_B)}{n_T n_F}\ \mathrm{[bit/s/Hz]},
\end{equation} 
where $M_B$ is the number of eMBB codewords in the codebook of each eMBB user. 
For URLLC users, we similarly define the rate $R_U$ as 
\begin{equation}
R_U = \frac{\log_2 (M_U)}{n_F}\ \mathrm{[bit/s/Hz]},
\end{equation}
where $M_U$ is the number of URLLC codewords in the codebook used by an URLLC user for each information packet. Furthermore, due to the limited blocklength transmission of the URLLC user, we explicitly define a constraint on its error probability $\mathrm{Pr}[E_U]$ as $\mathrm{Pr}[E_U] \leq\epsilon_U$. Finally, following the discussion above, we define as $L_U$ the access latency, i.e., the maximum number of minislots an URLLC user has to wait before transmitting a packet. Also relevant for URLLC users is the worst-case access latency $L_U$, as defined above.
\begin{figure}[ht]
	\centering
	\includegraphics[height= 6 cm, width= 16 cm]{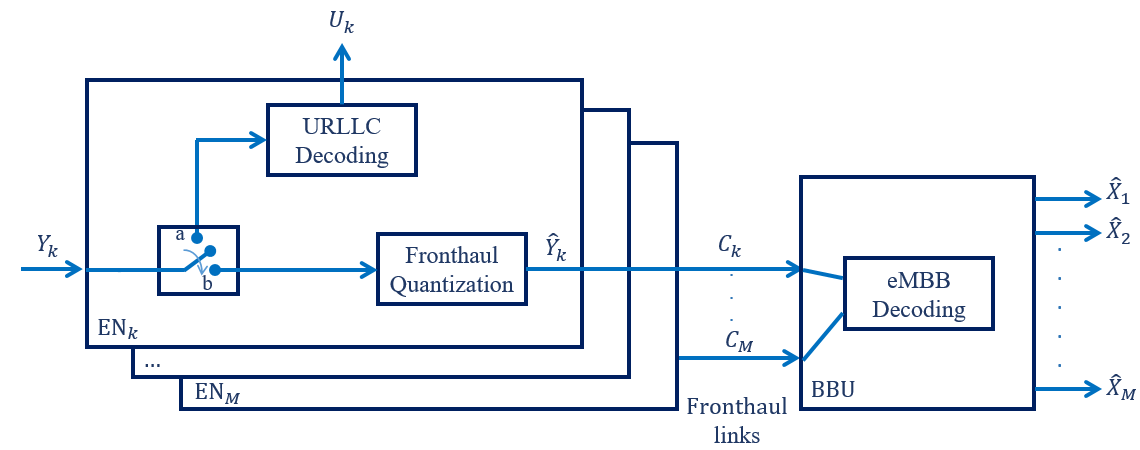}
	\caption{Block diagram representing the operation of ENs and BBU under OMA. The switch is in position a for minislots $t=L_U, 2L_U, ...$, while it is in position b otherwise.}
    \label{fig:internal_architecture_oma }
    \vspace{-2 mm}
\end{figure} 
\section{Orthogonal Multiple Access }

In this section, we consider the system performance in terms of eMBB rate $R_B$, URLLC rate $R_U$, and URLLC access latency $L_U$ for a fixed URLLC probability of error $\epsilon_U$ when assuming OMA. A block diagram representing the functionalities under OMA of both ENs and the BBU is shown in Fig. 3.

\subsection{URLLC Performance}
Due to latency constraints, URLLC packets are decoded at the local EN upon reception in the transmission minislot for $t=L_U,2L_U,...$. Under OMA, no interference is caused by the eMBB users on any active URLLC user. In order to obtain an achievable URLLC rate, we leverage the main result from \cite{polyanskiy2010channel}, which provides a finite blocklength characterization of the maximum achievable rate at a given probability of decoding error $\epsilon_{U}^{D}$. In fact, we recall that the URLLC blocklength $n_F$ is typically small and thus Shannon's asymptotic analysis cannot be applied. Accordingly, the URLLC rate can be well approximated as
\begin{equation}
R_{U}= \log(1+\beta^2 P_U) - \sqrt[]{\frac{V}{n_F}} Q^{-1}(\epsilon_U^{D}) \label{eq:RU},
\end{equation}
where $V$ is the channel dispersion
\begin{equation}
\begin{aligned}
V&= \frac{\beta^{2}P_U}{1 + \beta^{2}P_U}.
\end{aligned}
\end{equation}
\par 
Consider now the probability of error for an URLLC packet. This needs to account for two error events: (\textit{i}) more than one packet is generated by the URLLC user for each transmission opportunity and the given packet is not selected for transmission (blockage); and (\textit{ii}) the packet is transmitted but a decoding error occurs. Note that, in case (\textit{i}) of blockage, only one packet can be transmitted in the allocated minislot and the other packets cannot be delivered within the required worst-case delay of $L_U$ minislot. The probability of error for an URLLC transmission opportunity can hence be written as 
\begin{equation}
\mathrm{Pr}[E_U] = \sum_{n=1}^{L_U-1} p(n) \frac{n}{n+1} + \sum_{n=0}^{L_U-1} p(n) \frac{1}{n+1}\epsilon_{U}^{D}, \label{eq:proba}
\end{equation}
where $p(n)=\mathrm{Pr}[N_U(L_U)=n]$ is the distribution of the $N_{U}(L_U) \sim \mathrm{Bin}(L_U-1,q)$ binomial random variable representing the number of additional packets generated by the URLLC user during the remaining minislots between two transmission opportunities. The first term in \eqref{eq:proba} is the probability that a packet is lost due to blockage, which occurs if $n \geq 1$ additional packets are generated and the given packet is not selected. The second term in \ref{eq:proba} is the probability of case (\textit{ii}). Hence, the more stringent the requirement on the probability of error $\epsilon_U$ is, the more often there should be reserved opportunities for URLLC transmission, which in turn decreases the minislots available for eMBB.
\subsection{eMBB Rate}
Unlike delay-constrained URLLC traffic, eMBB messages are decoded jointly at the cloud in order to leverage the centralized interference management capabilities of the BBU. To this end, following the standard C-RAN operation, each $k$-th EN quantizes and compresses the received signal $Y_k$ for $t \neq L_U,2L_U, ...$ by using point-to-point compression (see \cite{parkandsimeone}\cite{simeone2012cooperative}), and forwards the resulting signal to the cloud over the fronthaul links as seen in Fig. 3. \par
Using \eqref{eq:1}, for each frequency $f$ and each minislot allocated to eMBB traffic, the quantized signal received at the BBU from $\mathrm{EN}_k$ can be written as
\begin{equation}
\hat{Y_{k}}^{f}=Y_{k}^{f}+Q_k^{f},
\end{equation}
where $Q_k^{f} \sim \mathcal{CN}(0,\sigma^{2}_{q})$ represents the quantization noise. 
From classical results in rate-distortion theory \cite{el2011network}\cite{parkandsimeone}, we have the following relationship between the quantization noise and the fronthaul capacity
\begin{equation}
\begin{aligned}
C&= \Big(1-\frac{1}{L_U}\Big) I(Y_k^{f}; \hat{Y}_k^{f})\\
&= \Big(1-\frac{1}{L_U}\Big) \log \Big( 1 + \frac{1 + \bar{P}_B (1+2\alpha^{2})}{\sigma_q^2}\Big), \label{eq:23}
\end{aligned}
\end{equation}
where $Y_{k}^{f}$ corresponds to the second case of equation \eqref{eq:1}, and 
\begin{equation}
\bar{P}_B= \frac{P_B}{1-L_U^{-1}}
\end{equation}
is the transmission power of the eMBB user under OMA.
The factor $(1-1/L_U)$ capture the fact that only a fraction $(1-1/L_U)$ of all minislots are occupied by eMBB transmissions. From \eqref{eq:23}, we obtain the quantization noise variance as
\begin{equation}
\sigma_q^2 = \frac{1+ (1+2\alpha^2)\bar{P}_B}{2^{\frac{C}{ (1-L_U^{-1})}}-1}. \label{eq:Q}
\end{equation}
\par In contrast with URLLC, the eMBB blocklength $n_T n_F$ is long enough to justify the use of standard asymptotic Shannon theory. To this end, considering the signals $ \hat{\mathbf{y}}^{f}=[\hat{Y}_{1}^{f} \cdots \hat{Y}_{2}^{f}]$ received from all $M$ ENs, the eMBB per user rate can be written as
\begin{equation}
\begin{aligned}
R_B&= \frac{(1-L_{U}^{-1})}{M} I(\mathbf{x}^{f};\mathbf{\hat{y}}^{f})\\
&= \frac{(1-L_U^{-1})}{M} \log \Bigg( \mathrm{det} (\mathbf{I} + \frac{\bar{P}_B}{1 + \sigma^{2}_{q}} \mathbf{H}\mathbf{H}^{\mathsf{T}} ) \Bigg)\\
&= \frac{(1-L_U^{-1})}{M} \sum_{m=0}^{M-1} \log \Bigg( 1+\Big(1 + 2\alpha \cos\Big(\frac{2\pi m}{M}\Big)\Big)^2 \frac{\bar{P}_B }{1+\sigma^{2}_{q}} \Bigg), \label{eq:sum}
\end{aligned}
\end{equation}
where the last equality follows from Szego's theorem \cite{gray1972asymptotic}. When $M \to \infty$ we can also write \eqref{eq:sum} in integral form \cite{simeone2012cooperative} as 
\begin{equation}
R_B= \Big(1-\frac{1}{L_U}\Big) \int_{0}^{1} \log \Big( \frac{(1 + 2\alpha \cos(2 \pi \theta))^2 \bar{P}_B + 1 + \sigma^{2}_q }{
1+\sigma^{2}_{q}} \Big) d\theta. \label{eq:RBortho}
\end{equation}

\section{Non Orthogonal Multiple Access}
In this section, we consider the performance of NOMA.  
\begin{figure}[ht]
	\centering
	\includegraphics[height= 8 cm, width= 16 cm]{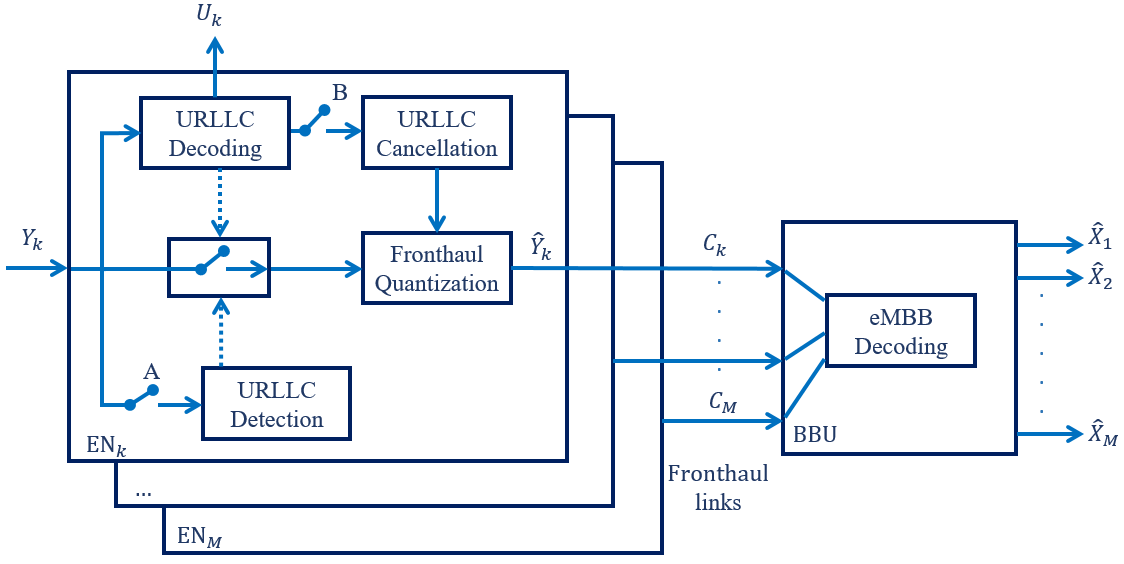}
	\caption{Block diagram representing the operation of ENs and BBU under NOMA. The boxed switch is always closed for the decoder that treats URLLC as noise, and is controlled by the indicated boxes for the other two decoders. Switch A is closed for the puncturing decoder, and switch B for the SIC decoder.}
    \vspace{-2 mm}
	\label{fig:internal_architecture }
\end{figure} 
\subsection{URLLC Performance}
With NOMA, as illustrated in Fig. 1(b), URLLC users transmit in any minislot in which a packet is generated with no additional access latency. We hence have the minimal access latency $L_U=1$. Furthermore, due to latency constraints, decoding of URLLC users cannot wait for the entire frame to be received and hence cannot benefit from interference cancellation of the eMBB signal. Therefore, the eMBB signal must be treated as interference when decoding URLLC signals at the EN.\par
In order to obtain an achievable rate for URLLC users in the presence of interference, we leverage {\cite[Theorem 2]{scarlett2017dispersion}}. This result provides a finite blocklength characterization of the maximum number of messages that can be transmitted on an interference channel under nearest neighbor decoding while treating interference as noise. Accordingly, the URLLC rate can be well approximated as
\begin{equation}
R_{U}= \log(1+S_U) - \sqrt[]{\frac{V}{n_F}} Q^{-1}(\epsilon_U^{D}), \label{eq:Ru}
\end{equation} where $S_U$ is the signal to interference plus noise ratio for the URLLC user
\begin{equation}
S_U= \frac{\beta^{2} P_U}{1 + (1+ 2\alpha^{2})P_B },
\end{equation}
with $(1+2\alpha^{2})P_B$ being the eMBB interference power, and the dispersion $V$ being given as
\begin{equation}
\begin{aligned}
V&= \frac{S_U}{1+S_U}
&= \frac{\beta^{2}P_U}{1 + (1+2\alpha^{2})P_B + \beta^{2}P_U}.
\end{aligned}
\end{equation}
\par With NOMA, an error can only occur when decoding fails, i.e., in case (\textit{ii}) listed above, and hence the probability of error is given as Pr[$E_U$] = $\epsilon_{U}^{D}$.
\subsection{eMBB Rate under Puncturing}

In this section, we consider a simple eMBB decoder that treats URLLC transmission as erasures. Accordingly, whenever a URLLC user is active in a given cell $k$ and minislot $t$, the signal $\mathbf{y}_{k}(t)$ received in the given minislot is discarded by $\mathrm{EN}_k$ and not forwarded to the BBU. This approach is under consideration within 3GPP. As we discuss next, in a C-RAN system, discarding symbols at the ENs has the effect of increasing the resolution of the samples that are quantized and conveyed to the BBU. In Fig.~\ref{fig:internal_architecture }, this scheme corresponds to the case where switch A is closed and switch B is open.\par
To elaborate, assume that each EN detects the transmissions of URLLC devices, and that it compresses and forwards only the signals received during minislots free of interference from URLLC transmission. Under this assumption, the signal $\tilde{Y}^{f}_{k}$ received at the cloud from $\mathrm{EN}_k$ on frequency $f$ can be written as
\begin{equation}
\begin{aligned}
\tilde{Y}_k^{f}=B_k(X_k^{f} + \alpha X_{k-1}^{f} + \alpha X_{k+1}^{f}) + Z_k^{f} + Q^{f}_{k} ,\label{eq:7}
\end{aligned}
\end{equation}
where the random variable $B_k=1-A_k\sim \mathcal{B}(1-q)$ indicates the absence ($B_k=1$) or presence ($B_k=0$) of URLLC transmissions in the given minislot. According to \eqref{eq:7}, the received signal $\tilde{Y}_k^{f}$ carries no information when the URLLC user is active ($B_k=0$). Otherwise, the signal contains the contributions of the eMBB users and of the quantization noise.
In matrix form, the signal in \eqref{eq:7} received across all ENs and frequencies can be equivalently written as
\begin{equation}
\mathbf{\tilde{Y} =BHX+ N + Q} \label{eq:8},
\end{equation}
with definitions given in Section II and with matrix $\mathbf{B}=\mathrm{diag}(B_1, B_2, \ldots,B_M)$.
\par In order to enable decoding, the BBU at the cloud needs to be informed not only of the signals \eqref{eq:8} for all the minislots with $B_k=1$, but also of the location of such minislots. To this end, each $\mathrm{EN}_k$ collects an i.i.d. binary vector containing the $n_T$ Bernoulli variables $B_k \sim \mathcal{B}(1-q)$. The number of bits needed to be communicated from $\mathrm{EN}_k$ to the BBU in order to ensure the lossless reconstruction is given by $n_T H(q) \mathrm{bits},$ where $H(q) = -q\log q - (1-q)\log(1-q)$ is the binary entropy function \cite{el2011network}.
\par
Based on the discussion above, imposing fronthaul capacity constraint yields the condition 
\begin{equation}
n_F n_T C = n_F n_T (1-q)I(Y_{k}^{f};\tilde{Y}_{k}^{f}| B_k = 1) + n_T H(q) \label{eq:totalC},
\end{equation}
where $Y_{k}^{f}=\tilde{Y}_{k}^{f}-Q^{f}_{k}$. The first term in \eqref{eq:totalC} represent the overhead needed to communicate the quantized signals \eqref{eq:7}.
From \eqref{eq:totalC}, the quantization noise is given by
\begin{equation}
\sigma^{2}_{q}= \frac{1 + (1+2\alpha^2)P_B }{2^{\frac{n_F C - H(q)}{n_F (1-q)}}-1} \label{eq:sigmaq}.
\end{equation}
\par The per-user rate is given by the mutual information
\begin{equation}
\begin{aligned}
R_B&=\frac{1}{n_F M} I(\mathbf{X};\tilde{\mathbf{Y}}|\mathbf{B})\\
&= \frac{1}{M} \mathbb{E}_{\mathbf{B}} \Big[\log \mathrm{det}\Big(\mathbf{I}+ S_B \mathbf{BHH^{\mathsf{T}}B}\Big) \Big] ,\label{eq:Rbexact}
\end{aligned}
\end{equation}
where the expectation is over the random matrix $\mathbf{B}$ and 
\begin{equation}
S_B= \frac{P_B}{1+\sigma_q^2}
\end{equation}
is the signal-to-noise ratio. The rate in \eqref{eq:Rbexact} can be easily evaluated numerically for any sufficient small value of $M$ and can be generally approximated via Monte Carlo's estimation. Furthermore, for $M \to \infty$, it can be computed exactly by using {\cite[Theorem 3]{tulino2007gaussian}} as
\begin{equation}
R_B=\int_{0}^{1-q} \log(1+f(y,S_B))dy \label{eq:RbI0},
\end{equation}
where $f(y,S_B)$ is the solution of the non-linear equation
\begin{equation}
\begin{aligned}
&\frac{f(y,S_B)}{1+f(y,S_B)}=\\
&\int_{-\frac{1}{2}}^{\frac{1}{2}} \frac{ \Phi(\theta)-1 - \sigma^{2}_q}{1+y(\Phi(\theta)-1 -\sigma^{2}_q)+ (1-y)f(y,S_B)} d\theta ,\label{eq:I0}
\end{aligned}
\end{equation}
with function \begin{equation}
\Phi(\theta)= (1+2\alpha \mathrm{cos}(2\pi\theta))^{2}\frac{P_B}{1+\sigma_q^2} + 1 + \sigma_q^{2}.
\end{equation}
\subsection{eMBB Rate by Treating URLLC As Noise } 
We now study the case in which the EN does not discard the signals received in each minislot if the URLLC user is active. In contrast, all received signals are quantized and forwarded to the BBU.The BBU decodes the eMBB messages while treating URLLC signals as noise. This decoding scheme corresponds to the case in Fig.~\ref{fig:internal_architecture } where switches A and B are open, and the boxed switch is always closed.  \par 
In this case, using \eqref{eq:Y_noma}, the quantized signal sent by $\mathrm{EN}_k$ to the BBU can be written as 
\begin{equation}
\hat{Y}_{k}^{f}= Y_{k}^{f} + Q_{k}^{f},
\end{equation}
where $Y^{f}_{k}$ is defined in \eqref{eq:Y_noma}. Samples with active URLLC devices can be compressed separately from those in which the device is not active. Imposing the fronthaul capacity constraint thus yields the following condition
\begin{equation}
\begin{aligned}
C &= I(Y_{k}^{f}; \hat{Y}_{k}^{f}| A_k)=  q \log\Big(1 + \frac{1+ \beta^2 P_U+ (1+2\alpha^2)P_B}{\sigma^{2}_{q}} \Big) \\ 
&+ (1-q) \log\Big(1 + \frac{1+ (1+2\alpha^2)P_B}{\sigma^{2}_{q}} \Big) ,\label{eq:CQ}
\end{aligned}
\end{equation}
where the first term corresponds to the minislot where the URLLC user is active ($A_k=1$, with probability $q$) and the second term for the case where the URLLC user is not active ($A_k=0$ with probability $1-q$). The quantization noise power $\sigma^{2}_{q}$, which is assumed to be the same for both classes of samples, can be computed by solving \eqref{eq:CQ} numerically. \par
Assuming that the BBU can detect when an URLLC user is active in each cell, the per-user sum-rate can be written as the mutual information
\begin{equation}
\begin{aligned}
R_{B}&= \frac{1}{Mn_F} I(\mathbf{X};\hat{\mathbf{Y}}|\mathbf{A}) \\
&= \frac{1}{M} \mathbb{E}_{\mathbf{A}} \bigg[ \log \frac{\mathrm{det}((1+\sigma^{2}_{q})\mathbf{I}+P_{B}\mathbf{HH}^{\mathsf{T}} + \beta^{2}P_U \mathbf{A})}{\mathrm{det}((1+\sigma^{2}_{q})\mathbf{I}+\beta^{2}P_U \mathbf{A})}\bigg]\\ \label{eq:28}
\end{aligned},
\end{equation}
where the average is taken over all possible values of the random matrix $\mathbf{A}$. Equation \eqref{eq:28} can be evaluated as \eqref{eq:Rbexact} using numerical methods.
\subsection{eMBB Rate via Successive Interference Cancellation (SIC) of URLLC}
We now study a more complex receiver architecture, whereby SIC of URLLC packets is carried out at the ENs prior to fronthaul quantization. More specifically, if an URLLC user is active and its message is decoded correctly at the receiving EN, the URLLC message is canceled by the EN. If decoding is unsuccessful, the URLLC message is instead treated as an erasure. \par
With this scheme, each EN quantizes the received signals only for the minislots that are either free of URLLC transmissions or that contain URLLC messages that were successfully decoded and canceled at the EN. This scheme corresponds to the case in Fig.~\ref{fig:internal_architecture } where switch B is closed and switch A is open.
As a result, the received signal at the BBU from $\mathrm{EN}_k$ can be written as \eqref{eq:7} but with an erasure probability of $q\epsilon_{U}^{D}$ instead of $q$. This is because an erasure occurs if the URLLC user is active and decoded incorrectly. As a result, the eMBB rate can be evaluated as \eqref{eq:Rbexact} and \eqref{eq:RbI0} under the mentioned substitution of the erasure probabilities.
\section{Numerical Results and Discussion}
In this section, we provide numerical results to bring insights, based on the analysis developed in the previous sections, on the achievable performance trade-offs between eMBB and URLLC traffic types under both OMA and NOMA and on the impact of key system parameters such as fronthaul capacity. We set $n_F=10$ subcarriers,\ $P_B= 5\ $dB$,\ P_U=10\ $dB, $\beta=1$, and $\epsilon_U=10^{-3}$.\par

\begin{figure}[t]
	\centering
	\includegraphics[height= 7 cm, width= 10 cm]{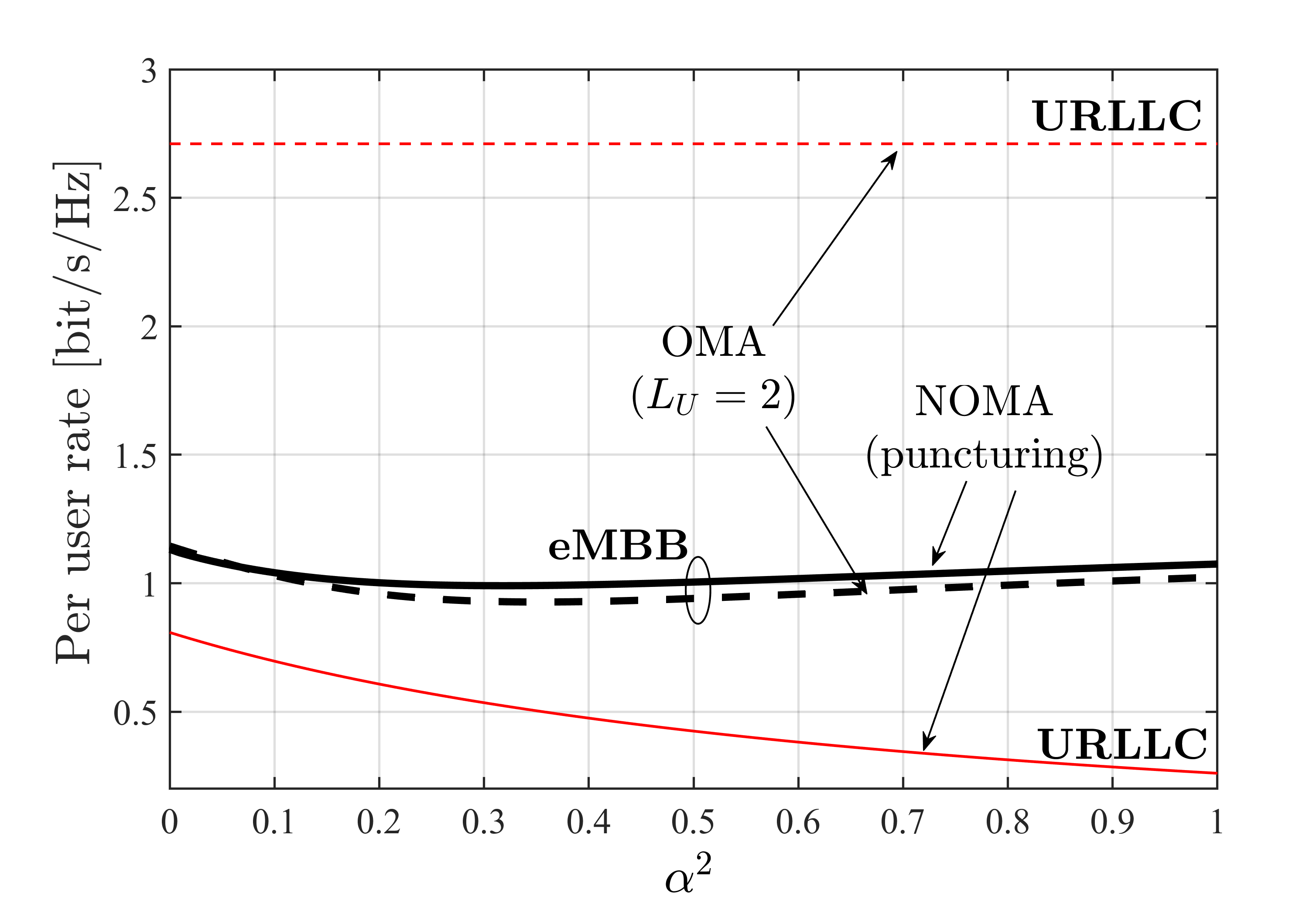}
	\caption{eMBB and URLLC per-user rates as a function of the inter-cell power gain $\alpha^{2}$ under OMA with $L_U=2$ and NOMA with puncturing ($q = 0.001, \epsilon_{U}=0.001,\ C=1.5$).}
	\label{fig:plot1}
\end{figure}

To start, in Fig.~\ref{fig:plot1}, we plot the per-user eMBB and URLLC rates for both OMA and NOMA as a function of the inter-cell power gain $\alpha^{2}$ for $q=0.01$, and $C=1.5$. For OMA, we set a worst-case access latency for URLLC users of $L_U=3$ minislots. For NOMA, we consider here the simplest form of processing, namely puncturing studied in Section IV.B.
We observe that, in the given scenario with small $q$, OMA offers a higher URLLC transmission rate due to the absence of interference from eMBB users, but this comes at the price of the higher URLLC access latency $L_U=2$. In contrast, NOMA provides the minimal access latency of $L_U=1$, while supporting a lower URLLC rate that decreases as a function of the inter-cell interference $\alpha^2$ due to eMBB interference. Furthermore, for eMBB traffic, NOMA provides a larger rate due to the larger number of available minislots. Finally, under both NOMA and OMA, the eMBB rate first decreases as a function of $\alpha$ due to the increased inter-cell interference while benefiting from larger values of $\alpha$, thanks to the joint decoding carried out at the BBU. \par
\begin{figure}[ht]
	\centering
	\includegraphics[height= 7 cm, width= 10.4 cm]{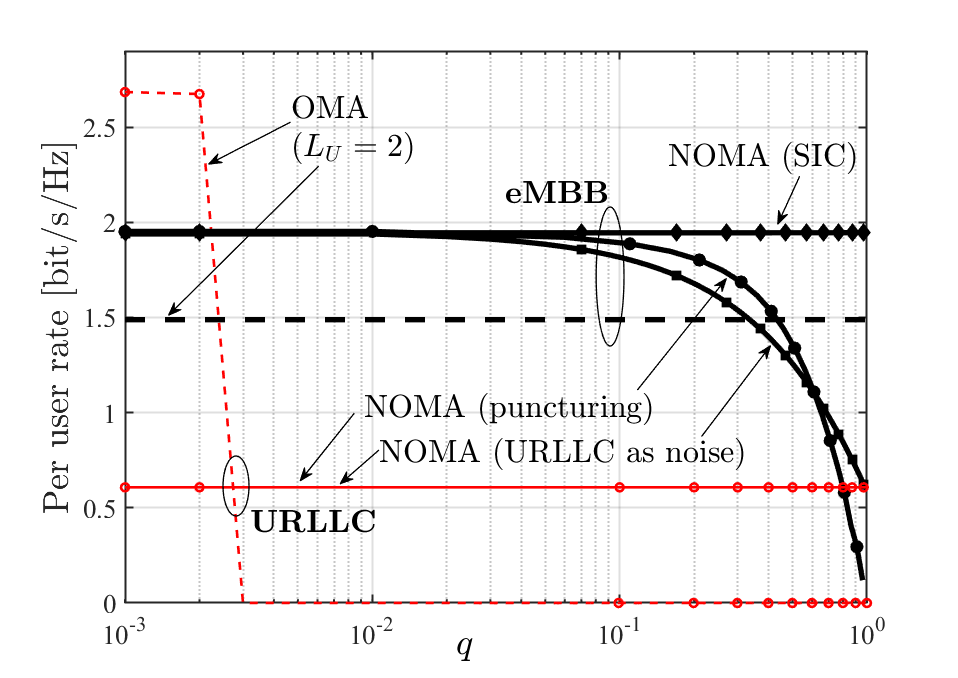}
	\caption{eMBB and URLLC per-user rates under OMA with $L_U=2$ and NOMA for different decoding strategies as function of $q$ ($\alpha^{2}=0.2,C=4$). }
    \vspace{-2 mm}
	\label{fig:plot4 }
\end{figure}

In Fig.~\ref{fig:plot4 }, we further investigate the per-user eMBB rate as a function of the URLLC traffic generation probability $q$ for $\alpha^2=0.2,C=4,$ and $L_U=2$ for OMA. The URLLC users' rate under OMA is seen to decrease quickly as a function of $q$. This is because, as $q$ increases, the error probability in \eqref{eq:proba} becomes limited by the probability that an URLLC packet is blocked due to an insufficient number of transmission opportunities. For NOMA, the URLLC rate is instead not affected by $q$. As for eMBB, for small values of $q$, here $q \leq 0.6$, treating URLLC signals as noise achieves the worst eMBB rate among the NOMA schemes. In fact, in this regime, if the fronthaul capacity is small, it is preferable not to waste fronthaul resources by quantizing samples affected by URLLC interference. In contrast, for larger values of $q$, puncturing becomes the worst-performing NOMA strategy, since the achievable eMBB rate becomes limited by the small number of useful received signal samples forwarded to the BBU. Finally, the more complex SIC scheme always provides the largest per-user eMBB rate thanks to the high probability of cancellation of URLLC signals at the EN.\par

\begin{figure}[ht]
	\centering
	\includegraphics[height= 7 cm, width= 10 cm]{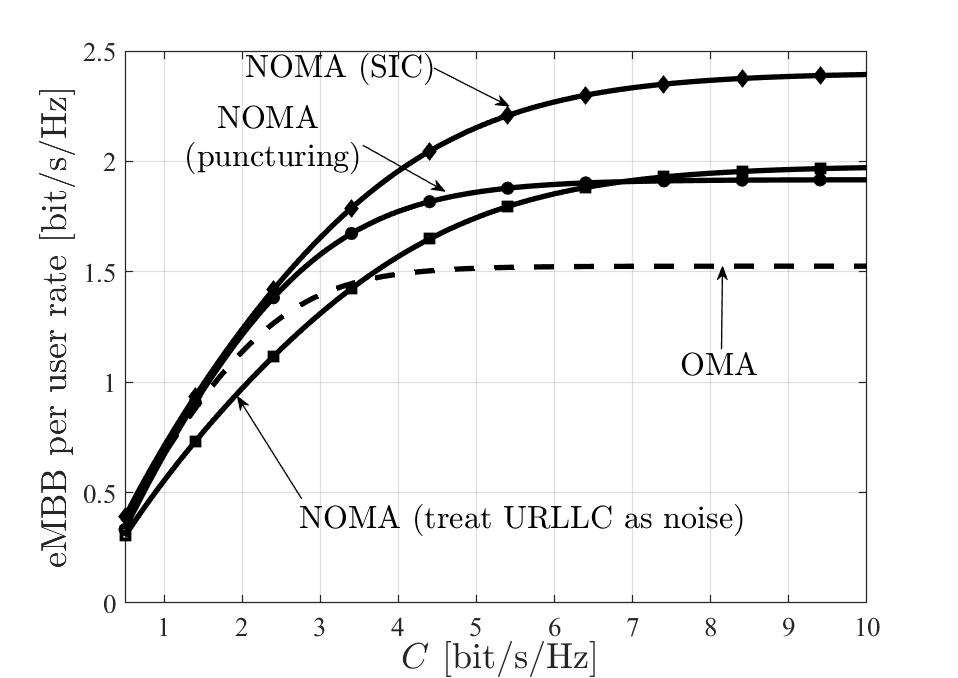}
	\caption{eMBB per-user rate under OMA with $L_U=2$ and NOMA for different decoding strategies as function of the fronthaul capacity $C$ ($\alpha^{2}=0.4,q=0.3$). }
	\label{fig:plot8 }
\end{figure}

In Fig.~\ref{fig:plot8 }, we plot the per-user eMBB rate as a function of the fronthaul capacity $C$ for $\alpha^2=0.4$ and $q=0.3$. We first note that, for small values of $C$, puncturing is preferable to treating URLLC as noise, since, as explained above, it avoids wasting the limited fronthaul resources on samples that are corrupted by URLLC interference. In this regime, puncturing provides the same performance as SIC, with the added benefit of a lower complexity and power consumption at the ENs. 
For larger fronthaul capacities, the quantization noise tends to zero, and thus treating URLLC as noise outperforms puncturing, given that it allows the BBU to make full use of the received signals. Moreover, NOMA with SIC provides the largest rate.
Finally, both Fig.~\ref{fig:plot4 } and Fig.~\ref{fig:plot8 } indicate that, with a sufficiently powerful decoder, such as SIC, the eMBB rate can be improved under NOMA as compared to OMA.\par
\begin{figure}[t]
	\centering
	\includegraphics[height= 7 cm, width= 10 cm]{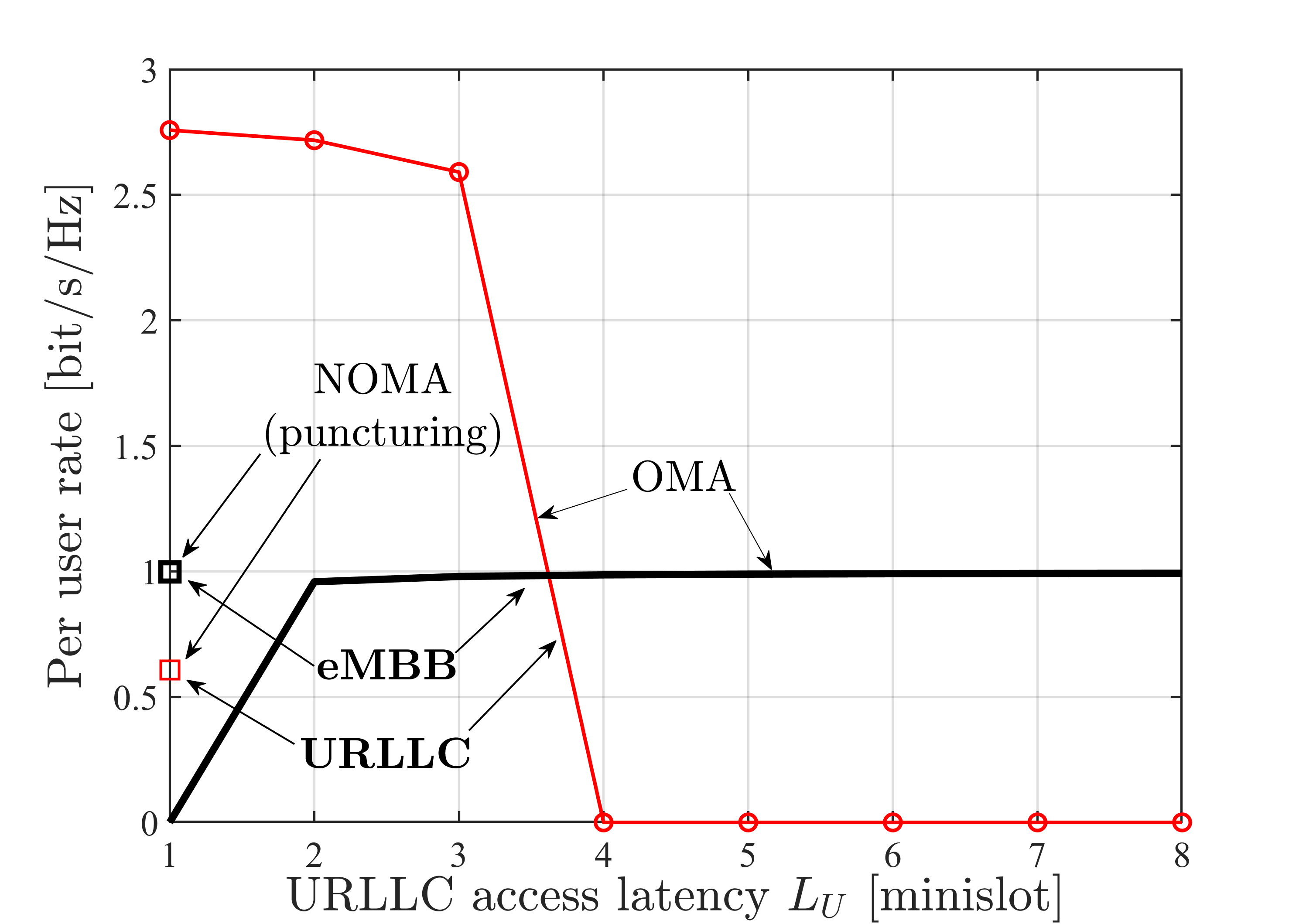}
	\caption{eMBB and URLLC per-user rates as a function of the URLLC access latency $L_U$ under OMA and NOMA with puncturing ($q=0.001, \alpha=0.2, C=1.5$).}
	\label{fig:plot2 }
\end{figure}

In Fig.~\ref{fig:plot2 }, we study the trade-off between the eMBB and URLLC per-user rates as a function of the access latency $L_U$. We set $\alpha=0.2$, $q=0.0.1$, $\epsilon_U=10^{-3}$ and $C=1.5$. Under OMA, the URLLC per-user rate decreases when the access latency $L_U$ grows due to the increased probability of URLLC packet blockage. To compensate to this contribution to the probability of error in \eqref{eq:proba}, one needs to reduce the probability of decoding error $\epsilon_{U}^{D}$, causing the rate to decrease (see \eqref{eq:RU}). In contrast to OMA, NOMA provides minimal and constant URLLC latency equal to $L_U=1$, but at the price of a lower rate due to interference from eMBB transmission. In addition, NOMA provides an eMBB rate comparable to OMA.

\section{Conclusions}
This work has investigated for the first time the performance trade-offs between eMBB and URLLC traffic types in a multi-cell C-RAN architecture under OMA and NOMA access strategies. As pointed out in \cite{Anand2017JointSO}\cite{networkslicingpopvskiandsimeone}, OMA offers eMBB users interference-free minislots, but it decreases the total number of symbols available for transmission and hence potentially the eMBB spectral efficiency. We argued that, while reducing mutual eMBB-URLLC interference, OMA was seen to have the disadvantage for URLLC users of introducing errors caused by packet drops due to an insufficient number of allocated transmission opportunities. As for NOMA, we have highlighted the significant gains accrued by SIC of URLLC traffic at the edge thanks to the high reliability requirements of URLLC. We have also revealed the potential benefits of puncturing in improving the efficiency of fronthaul usage by discarding  received minislots affected by URLLC interference. 
\bibliographystyle{IEEEtran}
\bibliography{Biblio}

\end{document}